\documentclass[reprint, prb, aps, groupedaddress, twocolumn, showkeys]{revtex4-2}

\usepackage{amssymb}
\usepackage{amsmath}
\usepackage{graphicx}
\usepackage{bm}
\usepackage{dcolumn}
\usepackage{color}
\usepackage{ulem}
\usepackage{float}
\usepackage{textcomp}
\restylefloat{table}
\usepackage{verbatim}
\usepackage{epstopdf}
\usepackage[colorlinks]{hyperref}
\usepackage[mediumspace,mediumqspace,squaren]{SIunits}
\usepackage[utf8]{inputenc}

\definecolor{mygreen}{RGB}{0, 88, 39}
\definecolor{myblue}{RGB}{46, 48, 146}

\hypersetup{
    colorlinks=true,
    linkcolor=blue,
    filecolor=magenta,      
    urlcolor=blue,
    citecolor=blue
}

\usepackage{pifont}

\usepackage{titlesec, titletoc}
\titleformat*{\section}{\boldmath\bfseries}
\titleformat*{\subsection}{\boldmath\bfseries}

\let\oldmaketitle\maketitle
\renewcommand\maketitle{{\bfseries\boldmath\oldmaketitle}}

\begin{document}

\title{Developing fractional quantum Hall states at $\nu$ = $\dfrac{1}{7}$ and $\dfrac{2}{11}$ in the presence of significant Landau level mixing}

\author{Siddharth Kumar\ Singh}
\author{A.\ Gupta}
\author{P. T.\ Madathil}
\author{C.\ Wang}
\author{K. W.\ Baldwin}
\author{L. N.\ Pfeiffer}
\author{M.\ Shayegan}
\affiliation{Department of Electrical and Computer Engineering, Princeton University, Princeton, New Jersey 08544, USA}

\begin{abstract}
Termination of the fractional quantum Hall states (FQHSs) and the emergence of Wigner crystal phases at very small Landau level filling factors ($\nu$) have been of continued interest for decades. Recently, in ultra-high-quality, dilute GaAs 2D electron systems (2DESs), strong evidence was reported for FQHSs at {$\nu=1/7, 2/13$ and 2/11} which fall in the {$\nu = p/(6p\pm1)$} Jain series of FQHSs, interpreted as integer ({$p = 1$}, 2) QHSs of 6-flux composite fermions ({$^6$CFs}). These states are surrounded by strongly-insulating phases which are generally believed to be Wigner crystals. Here, we study an ultra-high-quality 2DES confined to an AlAs quantum well where the 2D electrons have a much larger effective mass ($m^*\simeq 0.45 m_e$) and a smaller dielectric constant ($\epsilon\simeq10\epsilon_0$) compared to GaAs 2D electrons ($m^*\simeq 0.067 m_e$ and $\epsilon\simeq13\epsilon_0$). This combination of $m^*$ and $\epsilon$ renders the Landau level mixing parameter $\kappa$, defined as the ratio of the Coulomb and cyclotron energies, $\simeq 9$ times larger in AlAs 2DESs ($\kappa\propto m^*/\epsilon$). Qualitatively similar to the GaAs 2DESs, we observe an insulating behavior reentrant around a strong $\nu=1/5$ FQHS, and extending to $\nu<1/5$.  Additionally, we observe a clear minimum in magnetoresistance at $\nu=2/11$, and an inflection point at $\nu=1/7$ which is very reminiscent of the first report of an emerging FQHS at $\nu=1/7$ in GaAs 2DESs. The data provide evidence for developing QHSs of $^6$CFs at very small fillings. This is very surprising because $\kappa$ near $\nu \simeq 1/6$ in our sample is very large ($\simeq4$), and larger $\kappa$ has the tendency to favor Wigner crystal states over FQHSs at small fillings. Our data should inspire calculations that accurately incorporate the role of Landau level mixing in competing many-body phases of $^6$CFs at extremely small fillings near $\nu=1/6$.
\end{abstract}

\maketitle

\section{Introduction}

The discovery of the fractional quantum Hall state (FQHS) at Landau level filling factor $\nu=1/3$ in a GaAs two-dimensional electron system (2DES) \cite{Tsui.PRL.1982} was a landmark in the study of phenomena arising from electron-electron interaction. This observation was soon explained by Laughlin \cite{Laughling.PRL.1983} who provided a wave-function for the FQHS at $\nu=1/3$, and predicted similar incompressible states at $\nu=1/5, 1/7, \ldots$. As the 2D electron mobility in GaAs was improved, a developing FQHS at $\nu=1/5$ was indeed identified by an inflection point in the longitudinal magnetoresistance ($R_{xx}$) \cite{Mendez.PRB.1983} in an insulating background, which soon turned into a strong FQHS with $R_{xx}$ approaching zero, and a quantized Hall resistance ($R_{xy}$) at low temperatures \cite{Mallett.PRB.1988, Sajoto.PRB.1990, Jiang.PRL.1990, Goldman.PRL.1993, Sajoto.PRL.1993, Willett.PRB.Termination.1988, Goldman.PRL.1990}. In the highest quality, low-density GaAs samples, besides deep minima in $R_{xx}$ at $\nu=1/3$ and 1/5, an inflection point was also observed at $\nu=1/7$ \cite{Goldman.PRL.1988}. These discoveries came as a surprise as it was long believed that when the kinetic energy of a 2DES is quenched at very small fillings ($\nu<<1$), the ground state should be a Wigner crystal (WC) phase \cite{Lam.PRB.1984, Levesque.PRB.1984}. Insulating phases observed between $2/9<\nu<1/5$ and for $\nu<1/5$ were interpreted as WCs pinned by disorder \cite{Jiang.PRL.1990, Shayegan.WC.review.1996}, based on their non-linear \textit{I-V} \cite{Goldman.PRL.1990, Jiang.PRB.1991}, noise characteristics \cite{Li.PRL.1991}, microwave resonance responses \cite{Andrei.PRL.1988, Chen.NatPhys.2006} and, more recently direct evidence in the form of bilayer commensurability oscillations \cite{Deng.PRL.2016}. Data obtained via other experimental techniques support this interpretation \cite{Kukushkin.PL.1991, Tiemann.NatPhys.2014, Jang.NatPhys.2017}. With yet more breakthroughs in GaAs 2DES quality, strong minima at $\nu=1/7, 2/13$ and 2/11 \cite{Pan.PRL.2002, Chung.PRL.2022, Chung.NM.2021} were observed in $R_{xx}$, suggesting that the FQHSs do not terminate for $\nu\lesssim1/6$. The appearance of the $R_{xx}$ FQHS minima superimposed on the highly-insulating background signals a close competition between the FQHS and WC phases in the very small $\nu$ regime \cite{Zuo.PRB.2020}.

\begin{figure*}[t!]
\includegraphics[width=2\columnwidth]{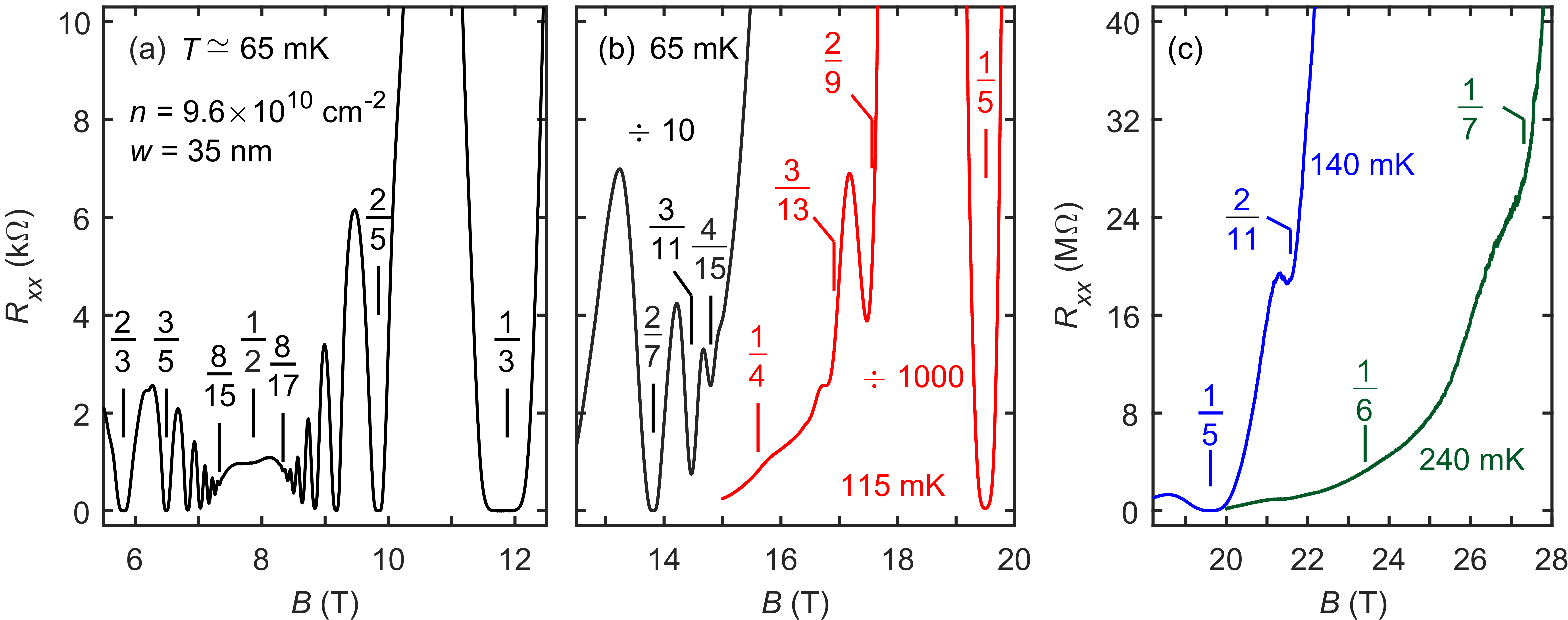}
\caption{\label{fig:fig1}$R_{xx}$ for 2D electrons confined to a 35-nm-wide AlAs QW as a function of perpendicular magnetic field, \textit{B}. Note the increasing $R_{xx}$ scales in panels (a)-(c), revealing the emergence of insulating phases at lower fillings. The $R_{xx}$ traces have been color coded based on the temperature at which they were taken, as indicated. (a) The ultra-high quality of our sample is evident from the numerous FQHSs flanking the $\nu=1/2$ CF Fermi sea. $R_{xx}$ minima following the $^2$CF Jain sequence at $\nu=p/(2p\pm1)$ can be seen up to $\nu=8/17$ and 8/15 on the high-field and low-field sides of $\nu=1/2$, respectively. (b) FQHSs around $\nu=1/4$ following the $^4$CFs sequence $\nu=p/(4p\pm1)$, and an insulating phase emerging past $\nu\simeq1/4$ can be seen in $R_{xx}$. {}(c) Developing FQHSs at very small fillings $\nu=1/7$ and 2/11 which fall in the $^6$CFs sequence of $\nu=p/(6p\pm1)$ can be observed as an inflection point and minimum in $R_{xx}$ for $p=1$ and 2, respectively. These states compete with the WC phases as evident from the highly-insulating background in which these FQHSs appear. Note that the LLM parameter, $\kappa$ is very large ($\simeq4$) for our sample near $\nu\simeq1/6$. Each panel displays the $R_{xx}$ trace at the lowest temperature at which reliable measurements could be made in the respective region in $\nu$.} 
\end{figure*}

Another crucial parameter that can modify the electron-electron interaction and therefore the ground states of 2D carrier systems is Landau level mixing (LLM). Denoted by $\kappa$, LLM is defined as the ratio of the Coulomb and cyclotron energies $\left(\kappa = \dfrac{e^2/4\pi\epsilon l_B}{\hbar eB/m^*}\propto m^*/\epsilon\right)$, where $\epsilon$ is the dielectric constant, $m^*$ is the effective mass, $B$ is the perpendicular magnetic field, and $l_B=\sqrt{\hbar/eB}$ is the magnetic length. Experimental evidence \cite{Santos.PRL.1992, Ma.PRL.2020, Maryenko.Nat.Comm.2018, Villegas.Rosales.PRR.2021, Santos.PRB.1992, Willett.PRB.1988, Shayegan.PRL.1990, Villegas.Rosales.PRL.2021, Csathy.PRL.2005} and theoretical calculations \cite{Yoshioka.JSoc.1984, Yoshioka.JSoc.1986, Price.PRL.1993, Zhao.PRL.2018} for the influence of $\kappa$ on the FQHS and WC phases have been reported before. In dilute GaAs 2D \textit{hole} samples, a reentrant insulating phase is observed for $2/5<\nu<1/3$ \cite{Santos.PRL.1992, Ma.PRL.2020, Santos.PRB.1992}; 2D \textit{electrons} in ZnO \cite{Maryenko.Nat.Comm.2018} also show a qualitatively similar phenomenon. The larger effective masses of GaAs 2D holes and ZnO 2D electrons lead to a more severe LLM (larger $\kappa$) in these systems, stabilizing the WC phase for $2/5<\nu<1/3$ and $\nu<1/3$. 2DESs confined to AlAs quantum wells (QWs) are yet another platform where remarkable strides in quality have been made \cite{Chung.PRM.2018} through the Al and Ga source purification \cite{Chung2.PRM.2018}, improved vacuum integrity, and optimization of modulation-doping to induce 2D carriers in AlAs QWs \cite{Chung.PRM.2017}. The improved samples have allowed the observation of new many-body states at zero magnetic field \cite{Hossain.PNAS.2020, Hossain.PRL.2021, Hossain.PRL.2022} and in the presence of high magnetic fields \cite{Hossain.PRL.2018, Chung.PRM.2018, Villegas.Rosales.PRR.2021, Hossain.PRL.2023}. Compared to GaAs 2DESs, 2D electrons in AlAs have a large effective band mass ($m^*\simeq0.45 m_e$) and small dielectric constant ($\epsilon\simeq10 \epsilon_0$), leading to much higher $\kappa$ ($\kappa_{AlAs}\simeq9\times\kappa_{GaAs}$). In this report, we focus on the termination of the FQHSs in an ultra-high-quality AlAs 2DES at very small fillings under the influence of significant LLM.

\section{Experimental details}

Our sample is a 35-nm-wide AlAs QW and has a 2D electron density of $n=9.6\times10^{10}$ cm$^{-2}$. The carriers are induced in the QW by modulation doping it symmetrically with Si on its two sides. The mobility of our sample is $0.8\times10^6$ cm$^2$/Vs measured at \textit{T} = 0.3 K. The 2D electrons in a 35-nm-wide AlAs QW occupy two degenerate valleys (\textit{X} and \textit{Y}) which have anisotropic effective masses, longitudinal mass $m_l = 1.0 m_e$ and transverse mass $m_t=0.20 m_e$, making the relevant electron mass in our system $m^*=\sqrt{m_l m_t}=0.45 m_e$. The degeneracy of the \textit{X} and \textit{Y} valleys can be lifted by uniaxial in-plane strain \cite{Hossain.PNAS.2020, Hossain.PRL.2022, Hossain.PRL.2023, Hossain.PRL.2021, Hossain.PRL.2018, Shayegan.Review.2006, Gunawan.PRL.2006}. In our sample, we do not apply any intentional strain, but the residual strain induced because of mounting the sample on the low-temperature sample holder likely breaks this symmetry so that only one valley is occupied, especially at very small $\nu$ that are relevant to the measurements presented here \cite{Lay.APL.1993, Poortere.APL.2002}. Contacts to the 2DES are made by alloying eutectic In:Sn to the QW in a reducing environment. The sample has van der Pauw geometry and all measurements are made using standard, low-frequency, lock-in techniques, in a dilution refrigerator. 

\section{Magnetotransport measurement data}

Figure \ref{fig:fig1}(a) depicts $R_{xx}$ as a function of $B$ at $T\simeq65$ mK. The sample's high quality is immediately evident from the myriad of features associated with interaction-driven, many-body states of 2D electrons displayed in the $R_{xx}$ trace. Notably, focusing on $\nu<1$, we observe several FQHS $R_{xx}$ minima flanking $\nu=1/2$. These minima occur at $\nu=2/3, 3/5, ..., 8/15$ on the low-field side and at $1/3, 2/5, ..., 8/17$ on the high-field side of $\nu=1/2$. As $B$ is raised, yet another sequence of $R_{xx}$ minima at $\nu=2/7$, 3/11 and 4/15 is observed at $T\simeq65$ mK (black trace in Fig. \ref{fig:fig1}(b)).  As the filling factor is lowered below $\nu=1/4$, 2D electrons in AlAs become increasingly resistive, with $R_{xx}$ rising beyond 100 k$\Omega$ at $T\simeq65$ mK. This leads to technical challenges in making reliable resistance measurements: (1) the highest resistance that we can measure reliably is limited by the input impedance of the preamplifier, (2) the ohmic contacts to the sample become extremely resistive and do not allow us to inject current through the sample. As we raise the temperature, the sample and contacts become less resistive, allowing us to pass current into the sample and measure voltages at other contacts reliably. At $T\simeq115$ mK, the red $R_{xx}$ trace in Fig. \ref{fig:fig1}(b) exhibits clear minima at $\nu=1/5$ and 2/9, and a shoulder at 3/13.

Sequences of $R_{xx}$ minima around $\nu=1/2$ and 1/4 seen in Figs. \ref{fig:fig1}(a) and \ref{fig:fig1}(b) correspond to the FQHSs and can be understood as the \textit{integer} quantum Hall states (IQHSs) of composite fermions (CFs) \cite{Jain.Book.2007, Jain.PRL.1989}, occurring at $\nu=p/(mp\pm1)$, where $p=1, 2,...$ is the integer filling factor of CFs and $m=2, 4, ...$ is the number of flux quanta attached to an electron to form the respective CF. The sequence of states around $\nu=1/2$ can be explained as the IQHSs of 2-flux CFs (${^2}$CFs), and similarly, the sequence around $\nu=1/4$ as IQHSs of 4-flux CFs (${^4}$CFs). (${^4}$CFs). It is notable that the sequence of $R_{xx}$ minima for $\nu\lesssim1/4$ is observed riding on an insulating background.

\begin{figure}[t!]
\includegraphics[width=1\columnwidth]{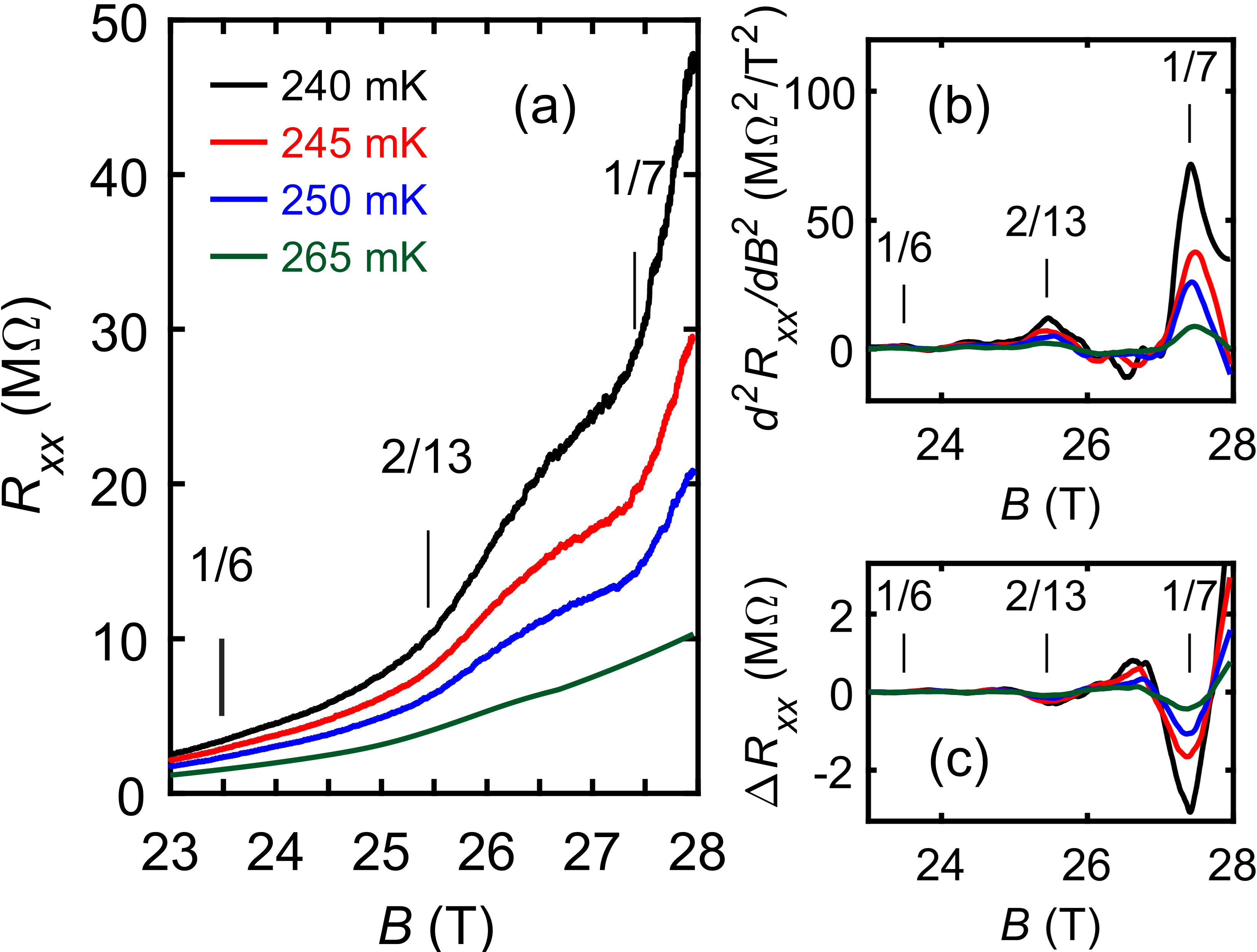}
\caption{\label{fig:fig3}  \color{black}(a) $R_{xx}$ vs. $B$ for $1/6\gtrsim\nu\gtrsim1/7$ at different temperatures. Vertical lines denote the expected positions of $\nu=1/7, 2/13$ and 1/6. The data show a clear inflection point in all four traces very close to the expected position of $\nu=1/7$. (b) Maxima in the second derivative of $R_{xx}$ are observed at $\nu=1/7$ and 2/13. (c) Data are shown after subtracting the background resistance from the $R_{xx}$. The background resistance is determined by filtering the $R_{xx}$ using a Savitzky-Golay filter with a 1.2 T window size. Clear minima are observed at $\nu=1/7$ and 2/13, and become weaker as the temperature is raised. The analyses in Figs. \ref{fig:fig3}(b) and (c) are indicative of a developing FQHS at $\nu=1/7$.}
\end{figure}

The major surprise in our data appears when we go to even higher magnetic fields and probe the region around $\nu=1/6$ which has not been studied before in a system with large $\kappa$. In Fig. \ref{fig:fig1}(c), we present the $R_{xx}$ trace at such low fillings where we observe a minimum at $\nu=2/11$ (blue $R_{xx}$ trace at 140 mK) and an inflection point at $\nu=1/7$ (green $R_{xx}$ trace at 230 mK). The expected positions of different $\nu$ are marked with vertical lines. The inflection point at $\nu=1/7$ in Fig. \ref{fig:fig1}(c) is reminiscent of the very first report of observation of a developing FQHS at $\nu=1/7$ \cite{Goldman.PRL.1988}. With improvements in sample quality, the inflection point at $\nu=1/7$ further developed \cite{Pan.PRL.2002}, finally culminating in a deep minimum in $R_{xx}$ \cite{Chung.PRL.2022}. Reference \cite{Chung.PRL.2022} reports a sequence of $R_{xx}$ minima at $\nu=1/7$, 2/13 and 2/11 which follow the Jain-sequence FQHSs at $\nu=p/(6p\pm1)$; these are the IQHSs of 6-flux composite fermions (${^6}$CFs). However, the LLM ($\kappa$) in GaAs for the parameters of Ref. \cite{Chung.PRL.2022} is $<1$. Our data provide evidence for the observation of developing FQHSs of ${^6}$CFs in an AlAs QW where $\kappa\simeq4$.

The evolution of $R_{xx}$ for $\nu<1/6$ at different temperatures is shown in Fig. \ref{fig:fig3}(a), with two noteworthy features. First, at such low fillings, $R_{xx}$ is of the order of several tens of M$\Omega$ at $T\simeq240$ mK and, as the temperature is raised, $R_{xx}$ decreases rapidly, by about a factor of five at the lowest fillings. This insulating behavior likely signals the formation of a magnetic-field-induced, pinned WC \cite{Sajoto.PRB.1990, Jiang.PRL.1990, Goldman.PRL.1993, Sajoto.PRL.1993, Goldman.PRL.1988, Shayegan.WC.review.1996, Goldman.PRL.1990, Jiang.PRB.1991, Li.PRL.1991, Andrei.PRL.1988, Chen.NatPhys.2006, Deng.PRL.2016, Kukushkin.PL.1991, Tiemann.NatPhys.2014, Jang.NatPhys.2017, Chung.NM.2021, Pan.PRL.2002, Chung.PRL.2022, Villegas.Rosales.PRR.2021, Willett.PRB.Termination.1988}. Second, the inflection point in $R_{xx}$ is reproducible and remains robust for $240\lesssim T\lesssim265$ mK. As mentioned before, previous observations of inflection points in $R_{xx}$ at $\nu=1/7$ \cite{Goldman.PRL.1988}, and the development of these inflection points to $R_{xx}$ minima riding on insulating phases in better quality samples \cite{Pan.PRL.2002, Chung.PRL.2022} have been attributed to the close competition between the FQHS and WC phases. 

{\color{black}The developing FQHS at $\nu=1/7$ is even more evident from the analyses described in Figs. \ref{fig:fig3}(b) and \ref{fig:fig3}(c). The inflection point in $R_{xx}$ vs. $B$ at $\nu=1/7$ appears as a maximum in its second derivative ($d^2R_{xx}/dB^2$) which is displayed in Fig. \ref{fig:fig3}(b) at different temperatures. As discussed earlier in the manuscript, the main contribution to $R_{xx}$ is from the insulating phase and appears as a rising resistance background. This resistance background (denoted as $\tilde R_{xx}$) is estimated by using a Savitzky-Golay filter with a 1.2 T window size on $R_{xx}$ vs. $B$ and subsequently subtracted from $R_{xx}$ to give $\Delta R_{xx} = R_{xx} - \tilde R_{xx}$. The resulting $\Delta R_{xx}$, which presumably is dominated by the contribution of the developing FQHSs, is displayed in Fig. \ref{fig:fig3}(c), and clearly shows a minimum at $\nu=1/7$. All three signatures of the developing FQHS at $\nu=1/7$ become weaker as the temperature is raised which is evident from Fig. \ref{fig:fig3}. Similar signatures of a developing FQHS at $\nu=2/13$ also appear in Figs. \ref{fig:fig3}(b) and \ref{fig:fig3}(c), indicating that both $d^2R_{xx}/dB^2$ and $\Delta R_{xx}$ are sensitive to the presence of competing FQHSs superimposed on insulating resistance backgrounds.}

\begin{figure}[t!]
\includegraphics[width=1\columnwidth]{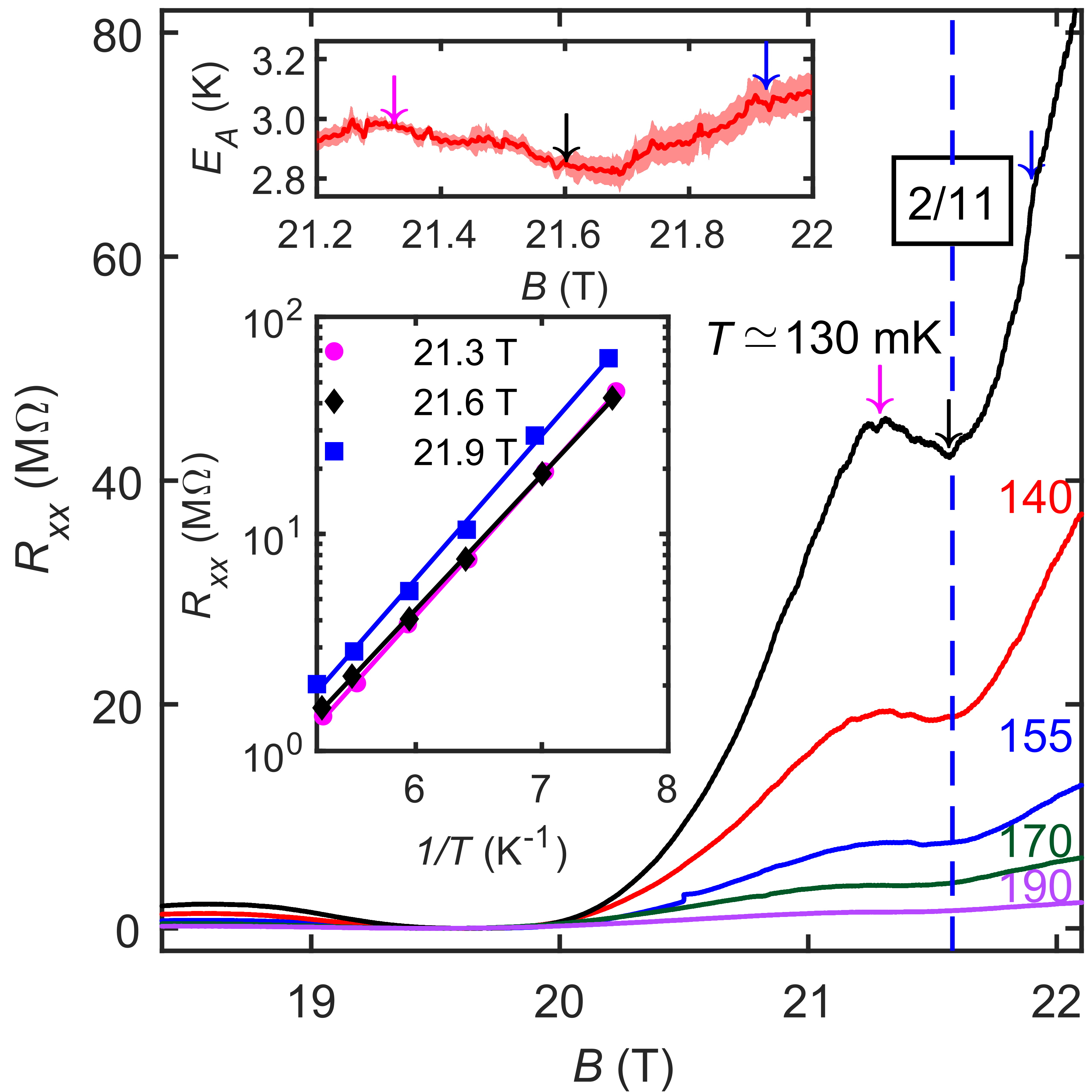}
\caption{\label{fig:fig4}Evolution of $R_{xx}$ around $\nu=2/11$ with temperature. The $R_{xx}$ minimum at $\nu=2/11$ is seen at low temperatures. \textit{Top inset}: Measured $E_A$ of the insulating phase for a small region around $\nu=2/11$. \textit{Bottom inset}: Arrhenius activation plots used to extract activation energies, $E_A$.}
\end{figure}

Similarly, we study the $R_{xx}$ minimum at $\nu=2/11$ for the temperature range $130\lesssim T\lesssim 190$ mK in Fig. \ref{fig:fig4}. The $R_{xx}$ minimum is seen up to $T\simeq170$ mK but becomes weaker at higher temperatures, as is typical of FQHSs. The insulating behavior of $R_{xx}$ is also at display around $\nu=2/11$. Using the temperature dependence of $R_{xx}$ around $\nu=2/11$, we have extracted the activation energy, $E_A$, by fitting $R_{xx}$ at a given $B$ to the expression $R_{xx}\propto e^{E_A/2kT}$, examples of which are shown in the lower inset in Fig. \ref{fig:fig4}. $E_A$ can provide a measure of the WC's defect formation energy \cite{Archer.PRB.2014, Esferjani.PRB.1990, Zhu.PRB.1995}. Values of $E_A$ at small fillings have been measured extensively before in GaAs 2D electrons, and have often been used to identify competing FQHSs in the WC regime \cite{Willett.PRB.Termination.1988, Jiang.PRL.1990, Mallett.PRB.1988, Chung.PRL.2022, Jiang.PRB.1991, Madathil.WCgaps.unpublished}. In the top inset of Fig. \ref{fig:fig4}, we show the magnetic field dependence of $E_A$ for our AlAs 2DES. It hovers around a value of $\simeq3$ K and displays a minimum very close to $\nu=2/11$, supporting the existence of a competing FQHS at $\nu=2/11$ \cite{Chung.PRL.2022}. The values of $E_A$ around $\nu\simeq2/11$ in our sample is comparable to the $E_A$ values in a GaAs 2DES at similar \textit{n} = $1.1\times10^{11}$ cm$^{-2}$, albeit in a wider, 50-nm-QW and with a mobility which is 30 times higher than that of our sample \cite{Madathil.WCgaps.unpublished}. However, one must be careful with a quantitative comparison because the value of $E_A$ for a WC depends strongly on factors such as LLM, finite-layer thickness and disorder, all of which are expected to lower $E_A$ down from its value for the pristine 2D WC. Also, the Fermi contour and effective mass are anisotropic in our AlAs 2DES while they are isotropic in GaAs 2DESs.

\begin{figure}[t!]
\includegraphics[width=1\columnwidth]{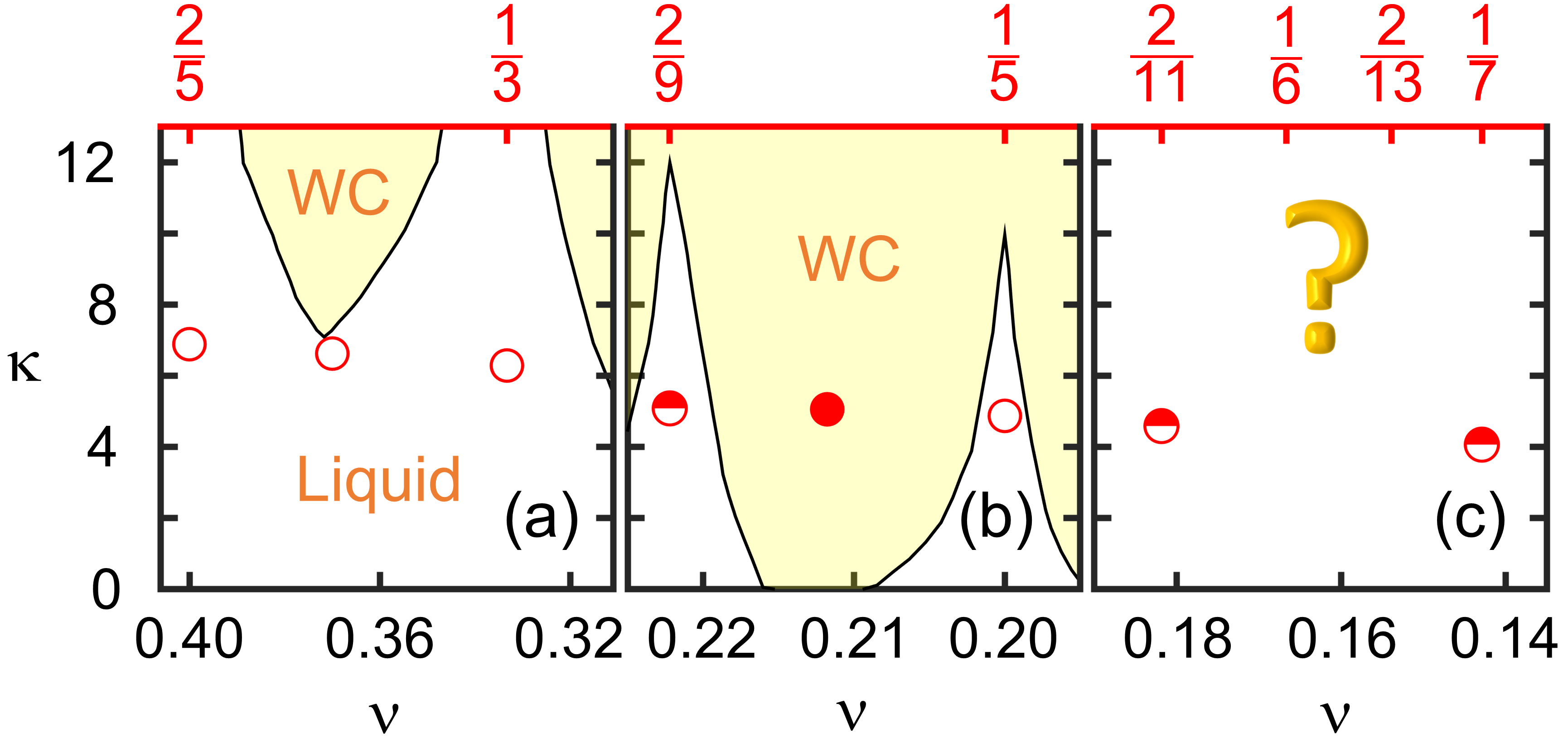}
\caption{\label{fig:fig2}(a,b) Calculated phase diagrams depicting the competition between the WC and the FQHSs under the influence of LLM as a function of $\nu$ \cite{Zhao.PRL.2018}. The WC phase is stable in the yellow regions of the phase diagrams. The experimental data points are denoted by circles. Open and solid red circles are used to denote the liquid and the WC phases, respectively, and half-filled circles denote FQHSs competing with insulating (WC) phases. For the magnitude of LLM ($\kappa$) present in our sample, the different phases agree reasonably well with the theoretical results. (c) The region of $\nu<1/5$ experimentally exhibits strongly insulating behavior, while features for competing FQHSs are seen at $\nu=1/7$ and 2/11. The yellow question mark highlights that no theoretical calculations akin to Figs. \ref{fig:fig2}(a) and \ref{fig:fig2}(b) exist around such small $\nu$.}
\end{figure}

\section{Discussion}

Figure \ref{fig:fig2} summarizes the significant role of LLM in determining the ground states of 2D electrons at small fillings. In Ref. \cite{Zhao.PRL.2018}, Zhao et al. quantitatively calculate and compare the ground-state energies of the WC and FQHSs for different values of $\kappa$ around $\nu = 1/3$ and 1/5. The resulting phase diagrams presenting the stability of the WC (yellow regions) and the correlated liquid phase (white regions) are shown in Figs. \ref{fig:fig2}(a) and \ref{fig:fig2}(b). Our experimental data, shown as red circles, agree well with the calculated phase diagram. As displayed in Fig. \ref{fig:fig2}(a), we observe strong FQHSs at $\nu=1/3$ and 2/5, and a liquid (non-insulating) phase between 2/5 and 1/3 (shown as open red circles), consistent with the theoretical results. For $1/5<\nu\lesssim2/9$, as shown in Fig. \ref{fig:fig1}(b), we observe an insulating behavior which we attribute to a WC phase (solid circle in Fig. \ref{fig:fig2}(b)) \cite{Villegas.Rosales.PRR.2021}, but at $\nu=1/5$ we observe a well-developed FQHS (open circle in Fig. \ref{fig:fig2}(b)). At $\nu=2/9$, we observe a deep minimum in $R_{xx}$, suggesting a developing FQHS, but the background is insulating. Therefore, we represent the state at 2/9 by a half-filled circle in Fig. \ref{fig:fig2}(b) to signify the close competition we observe experimentally between the FQHS and WC at this filling. Note that, in comparison, a very strong 2/9 FQHS is observed in very high quality GaAs 2DESs at a similar density, consistent with their much smaller LLM ($\kappa \simeq 0.56$ at $\nu=2/9$) \cite{Pan.PRB.2000}. 

Experimentally, for $\nu<1/5$, $R_{xx}$ exhibits a strongly insulating behavior indicative of pinned WC, while at $\nu=1/7$ and 2/11 there are developing FQHSs present in close competition with the WC phase; these are denoted by half-filled circles in Fig. \ref{fig:fig2}(c). Qualitatively, based on the theoretical calculations presented in Figs. \ref{fig:fig2}(a) and \ref{fig:fig2}(b) \cite{Zhao.PRL.2018}, we observe that: (1) The FQHSs at $\nu=1/3$ and 2/5 are quite robust and are the ground states even for $\kappa>18$ \cite{Zhao.PRL.2018}. In contrast, the FQHSs at $\nu=1/5$ and 2/9 make transitions to WC states for $\kappa >10$ and $> 12$, respectively. (2) The window in which the FQHSs are stable is narrower in $\nu$ at $\nu=1/5$ compared to $\nu=1/3$. Presumably, this window would be even narrower for FQHSs around $\nu=1/6$. Our data for a 2D system with significant LLM ($\kappa_{\nu=1/7}\simeq4.1$ and $\kappa_{\nu=2/11}\simeq4.6)$) compared to $\kappa<1$ in GaAs 2D electrons) evince that FQHSs at 1/7 and 2/11 are still quite competitive with WC phases even at very large $\kappa$. Thus they provide crucial data to test the validity and accuracy of future calculations, which we hope our data will stimulate.

\section{Summary}

While our data reported here highlight the stability of FQHSs in a 2DES with substantial LLM at very small filling factors, we remark that LLM has much broader implications. For example, numerous calculations have illuminated the importance of LLM in determining the exact ground-state wave function of the celebrated $\nu=5/2$ FQHS \cite{Rezayi.PRL.2017, Peterson.PRL.2014, Pakrouski.PRX.2015}. Also, recent 2D \textit{holes} in GaAs have shown new exotic \textit{even-denominator} FQHSs in the lowest LL \cite{Wang.PRL.2022, Wang.PNAS.2023, Wang.PRL.2023, holes.footnote, Gupta.PRM.2024}, an observation that has been explained by pairing of CFs induced by the severe LLM \cite{Zhao.PRL.2023}. Finally, the advent of exotic many-body physics in new 2D systems such as ZnO \cite{Maryenko.Nat.Comm.2018}, transition-metal dichalcogenides \cite{Shi.Nat.Nanotech.2020}, and bilayer graphene \cite{Tsui.Nature.2024} has brought forth yet more platforms with significant LLM. It remains to be seen what the carriers in these platforms exhibit as their quality is further improved by lowering the amount of disorder, so that the regime of very small filling factors could be reached.

\section{Acknowledments}

We acknowledge support by the U.S. Department of Energy Basic Energy (DOE) Sciences (Grant No. DEFG02-00-ER45841) for measurements, the National Science Foundation (NSF) (Grants No. DMR 2104771 and No. ECCS 1906253) for sample characterization, and the Eric and Wendy Schmidt Transformative Technology Fund for sample fabrication. The Princeton University portion of this research is funded in part by the Gordon and Betty Moore Foundation’s EPiQS Initiative, Grant GBMF9615.01 to Loren Pfeiffer.Our measurements were performed at the National High Magnetic Field Laboratory (NHMFL), which is supported by the NSF Cooperative Agreement No. DMR 2128556, by the State of Florida, and by the DOE. This research is funded in part by a QuantEmX grant from Institute for Complex Adaptive Matter (ICAM) and the Gordon and Betty Moore Foundation through Grant No. GBMF9616 to S. K. S., A. G., P. T. M., C. W. and M. S. We thank R. Nowell, G. Jones, A. Bangura and T. Murphy at NHMFL for technical assistance, and J. K. Jain and A. C. Balram for illuminating discussions.

\end{document}